\documentclass{PoS}
\usepackage{graphicx}
\usepackage{subfigure}
\usepackage{amssymb}
\usepackage{pifont}
\usepackage{cite}
\RequirePackage{longtable}

\title{POLAR measurements of the Crab pulsar}

\ShortTitle{POLAR Crab}

\author{Hancheng Li $^{ab}$, \speaker{Mingyu Ge} $^{b}$, Bobing Wu $^{b}$ on behalf of the POLAR collaboration\thanks{\emph{Abous us} at our website: polar.ihep.ac.cn or www.astro.unige.ch/polar\newline{} \hspace*{0.75cm}Cite as below: \newline{} \hspace*{0.75cm}Li, H. C., et al. POLAR measurements of the Crab pulsar. \emph{PoS (ICRC2017)}, 820 (2017), doi: 10.22323/1.301.0820} \\
\llap{$^a$}University of Chinese Academy of Sciences, Beijing 100049, China\\
\llap{$^b$}Key Laboratory for Particle Astrophysics, Institute of High Energy Physics, Beijing 100049, China\\
E-mail: \email{gemy@mail.ihep.ac.cn}}

\abstract{POLAR is a Compton polarimeter sensitive in the 50 to 500 keV energy range.
The Crab pulsar is a scientific target for POLAR on board the Chinese space
laboratory Tiangong-2 (TG-2). With its large Field of View (FoV), POLAR detected
significant pulsed signals from the Crab pulsar which is visible by POLAR
in about half of observation time. In this work, we present the preliminary
results including the pulse profile, timing and polarization measuring method.
First, we show the highly significant pulse profile observed by POLAR which
is compared to the results of other detectors including Fermi/LAT and INTEGRAL.
And the pulse profile as a function of theta incident angle and as a function of
channel number, which indicate that POLAR has a good detection performance,
have been showed. Second, we find that the timing of the Crab pulses are
accurately measured, which provides a unique verification and calibration to
the POLAR timing system. Finally, the potential polarization measurement of
the Crab pulsar is also discussed.
}

\FullConference{35th International Cosmic Ray Conference --- ICRC2017\\
		10--20 July, 2017\\
		Bexco, Busan, Korea}

\begin{document}

\section{Introduction}

The Crab pulsar is one of the most widely studied celestial objects which
was born in 1054, has a spin period of about 33\,ms, with a bright feature over
almost the full electromagnetic spectrum from radio to high energy $\gamma$-rays.
At all wavelengths this pulsar shows a double-pulse structure, with the
main pulse (P1) and the inter pulse (P2) separated by a phase of $144^{\circ}$  \cite{Eikenberry(1997)}\cite{Kuiper(2001)}\cite{Rots(1998)}\cite{
Molkov(2010)}\cite{Ge(2012)}. In addition, the polarization
of the Crab pulsar and Nebulae are also concerned since they have highly
magnetized filed. A sounding rocket mission \cite{Novick(1972)} reported
the first detection of X-ray polarization (5-20 keV) for the Crab nebula with a sounding
rocket payload. Then, the polarization fraction was measured with $19.2\pm1.0)$ at 2.6\,keV
by an instrument on-board the OSO-8 satellite mission \cite{Weisskopf(1976)}
during a 256 ks observation. More recently, the INTEGRAL satellite \cite{Forot(2008)} \cite{Dean(2008)},
AstroSAT and PoGO+ \cite{Chauvin(2017)} instruments have also measured the polarization of Crab emissions.

\begin{figure}
\begin{center}
\includegraphics[width= 0.9\textwidth, height = 10cm ]{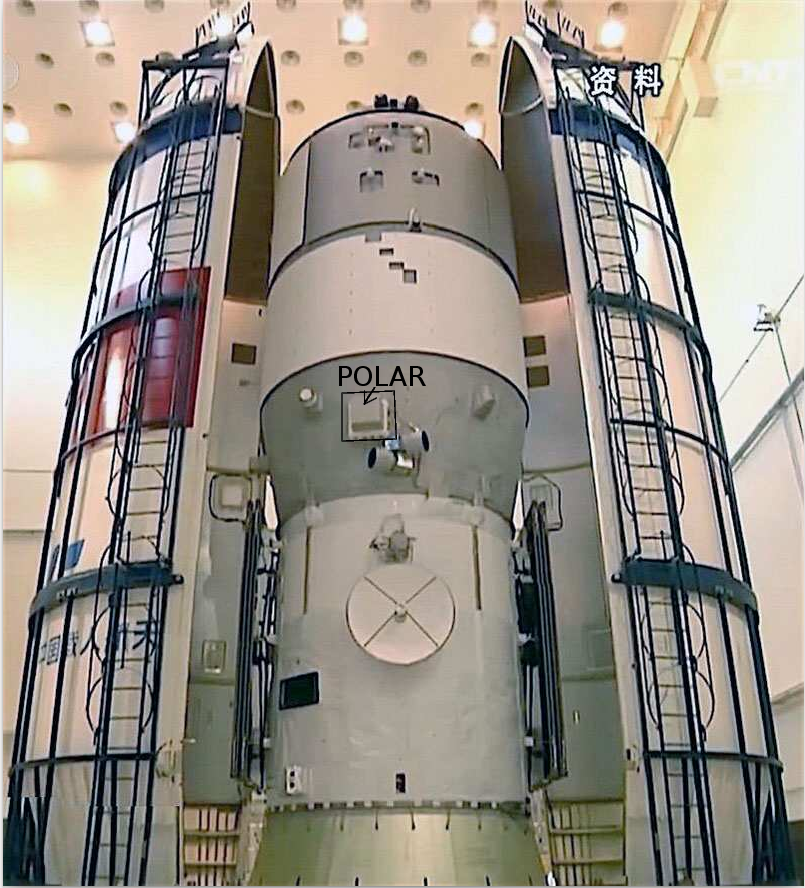}
\caption{The installation diagram of POLAR on TG-2. The outboard part is called obox, it's connected with ibox which is supporting electronics system inboard.
\label{fig1}}
\end{center}
\end{figure}

\begin{figure}[htbp]
\centering
\subfigure[]{
\begin{minipage}{7cm}
\centering
\includegraphics[width= 7 cm,height=7cm]{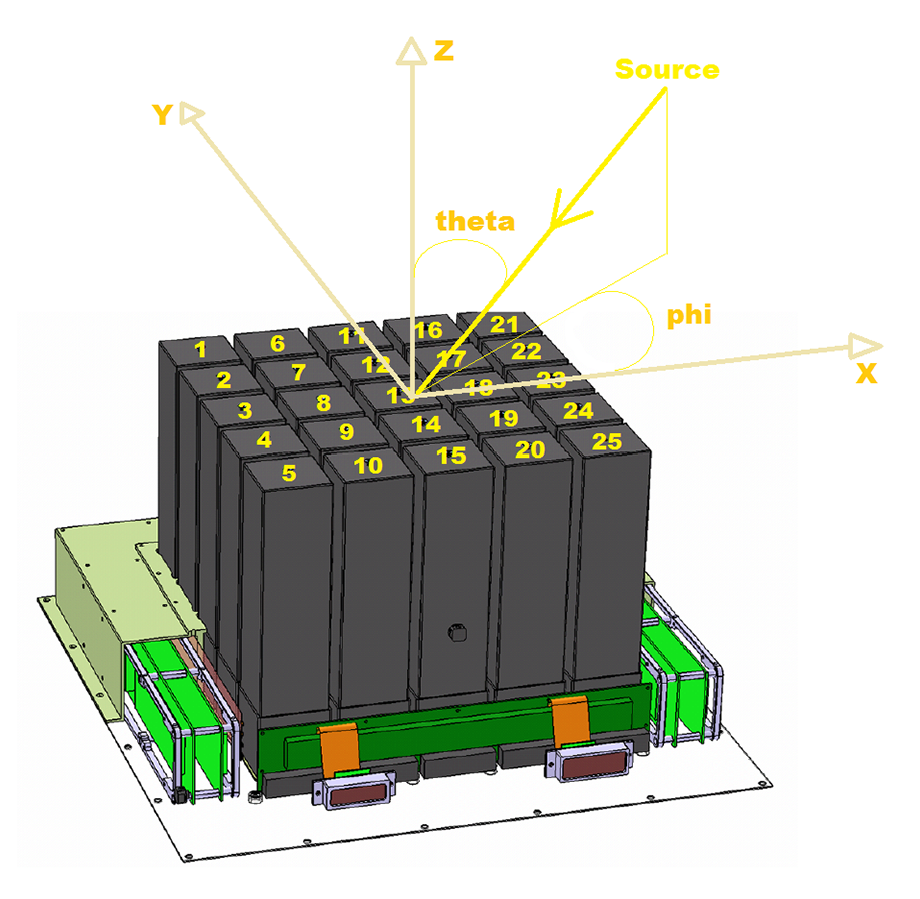}
\end{minipage}
}
\subfigure[]{
\begin{minipage}{7cm}
\centering
\includegraphics[width= 7 cm,height=7cm]{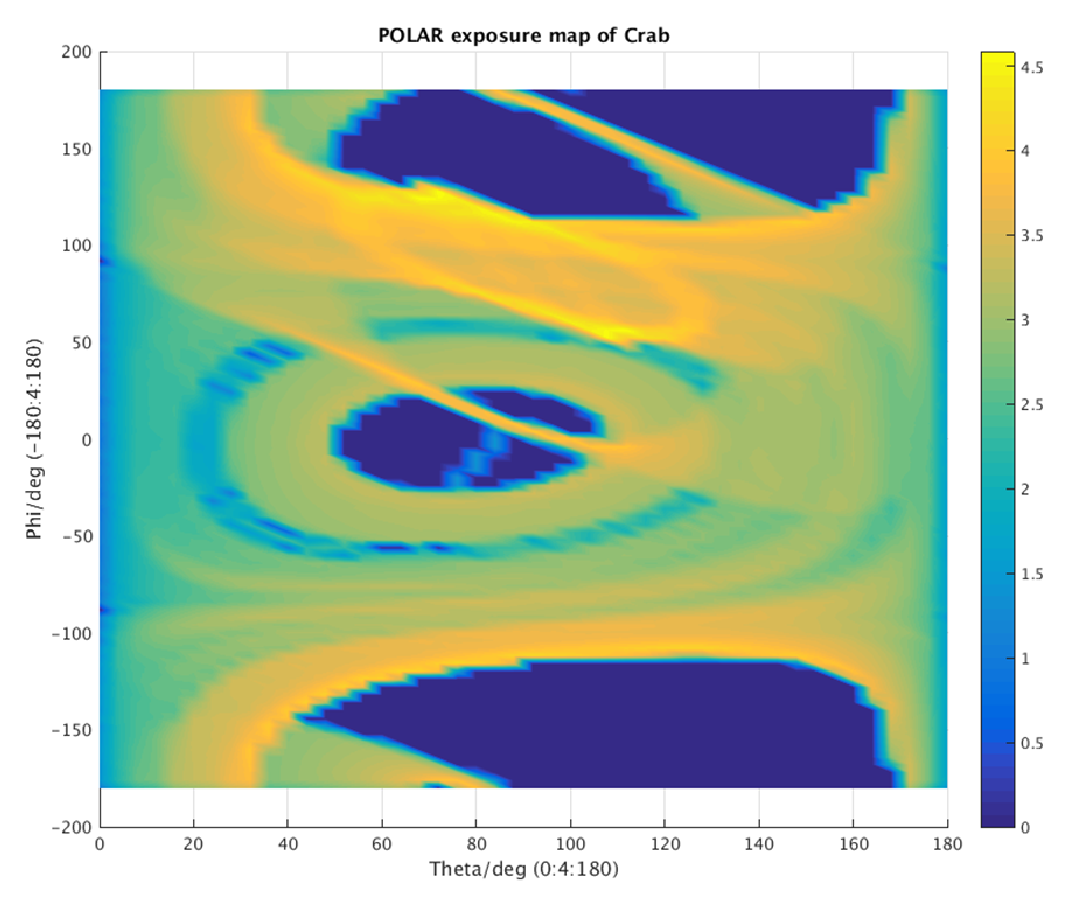}
\end{minipage}
}
\caption{(a) is the structure of POLAR detector. It consists of 25 modules, and each module has 8*8 detecting channels made of plastic scintillator bars. More working schematics of POLAR can be found in [12] and [13]. The incident angles are defined as theta and phi. Theta angle is the angle of Z-axis positive direction and the source vector from origin of coordinates to source point. Projecting the source vector onto X-Y plane, and phi is the angle between projection and X axis positive direction. (b) is the Crab exposure map on POLAR. It does not cover all incident directions.
\label{fig2}}
\end{figure}

POLAR started its mission on-board the Chinese space laboratory TG-2 after
the successful launch which was on 15th September, 2016 \cite{Merlin(2017)}.
The installation diagram of POLAR on TG-2 \footnote{http://www.collectspace.com/images/news-091516d-lg.jpg}
is shown in Figure \ref{fig1}. POLAR has a high opportunity to capture photons
from Crab pulsar thanks to its high sensitivity and large FoV when flying in-orbit,
even if POLAR has no pointing control system by itself. The structure of POLAR detector \cite{Sun(2012)} is shown in Figure \ref{fig2} (a).
When photons incident on POLAR, they tend to scatter in the detector. The deposited
energy of these scattering photons will be converted to digital signal, and if the
signal value exceed certain thresholds, through an complex process by electronic
system \cite{Produit(2005)}, these trigger information will be recorded.

The polarization of Crab pulsed emission can be measured by POLAR
since it is a polarimeter especially for GRB with an energy
range from 50 to 500\,keV. The basic concept is to measure
the mean degree of polarization and the azimuthal angle of the
polarization vector by analyzing the angular distribution of
the Compton scattering azimuthal angle of a sample of photons from GRB or other X-ray/ Gamma-ray emissions. The emission of the Crab pulsar
can also be observed because the pulsed photon can be accumulated
with the steady pulse phase considering the frequency evolution, though
it is very faint compared with the GRB flux in a short time interval.
However, it is very difficult to acquire the polarization directly
because the incident angle between the Crab and POLAR varies with time
and the signals could not be accumulated directly. Considering this
situation, some new methods should be developed together with
the Monte Carlo simulation.

\section{Observations and Data Reduction}

\subsection{Observations}

The Crab observation period for POLAR can obtained only through
the judgement of IN or OUT FoV. So we can calculate the incident
angle from Crab to POLAR detector by use of platform parameters
data (PPD) which is provided in every second. And we can get an exposure map of Crab observation by accumulating the same incident angle of observation together, as shown in Figure \ref{fig2} (b). Above all, in scientific observation
data (SCI), for each event, it has a lot of useful information,
such as trigger time, deposited energy signal, trigger position,
trigger number, etc. These information support us to perform the
following analysis.

\subsection{POLAR data reduction}

Crab photons are drown in a sea of background. Data reduction is necessary to
reduce background. First, the acquisited data lack of PPD should be rejected,
since we are not clear the incident angle without PPD, and we don't know
whether there is crab photon in SCI. Second, in order to ensure that crab
is within our visible region, the event whose theta incident angle larger than
100 degrees should be rejected. At last, in view of the scattering times of
Crab photons in POLAR detector rarely more than 5, we throw away the event whose
trigger number more than 5. The above reduction methods can improve ratio
of signal and noise for Crab analysis.

\subsection{Timing process}

We study the timing properties of the Crab pulsars with the
POLAR observations. First, the arrival time for each photon was converted
to the Solar System Barycentre using ephemeris DE405. For each
observation we obtained the period of a pulsar by folding
the observed counts to reach the maximum Pearson $\chi^2$.
Second, the pulse profile was folded with the spin frequency and
the time of arrival (TOA) was calculated from the peak phase of
the pulse and the reference time. Then timing parameters were solved
by the phase coherent timing method utilizing TEMPO2 \cite{Edwards(2006)} \cite{Hobbs(2006)}.
At the end, the Fermi-LAT \footnote{https://fermi.gsfc.nasa.gov/ssc/data/analysis/LAT\_essentials.html}
observations were analysed with the same process
to check whether the observation by POLAR was performed appropriately
especially for the timing system.

\section{Results from POLAR}

\subsection{Timing results}

\begin{figure}
\begin{center}
\includegraphics[width= 0.9\textwidth ,height = 10 cm ]{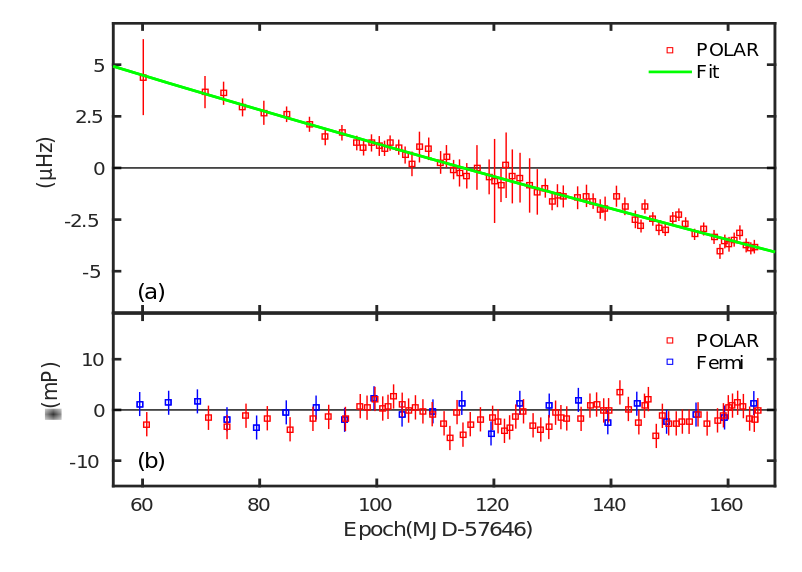}
\caption{Panel (a): the evolution of the spin frequency of the Crab pulsar observed by POLAR.
Each data point is subtracted by $(29.648422-t*3.68 \times 10^{-10} )$ to show the details
of its frequency evolution. The green line represents the fitted result. Panel (b): The time
residuals of the Crab pulsar observed by Fermi and POLAR, as represented by the blue and red
squares respectively.
\label{fig3}}
\end{center}
\end{figure}

From the almost consecutive observation from POLAR, we have
obtained the frequency evolution and timing results as shown
in Figure \ref{fig3} and Table \ref{table1}. First of all, the spin
frequency of the Crab pulsar is checked. Due to the large
background for the Crab pulsar, we combined the observation data
in every day to search the spin frequency of the pulsar. As shown
in Figure \ref{fig3} (a), the spin frequency of the pulsar decreases
with time significantly with time, which is consistent with the
ephemeris supplied by Jodrell Bank \cite{Lyne(1993)}
\footnote{http://www.jb.man.ac.uk/pulsar/crab.html}. Then,
TOAs were calculated with the spin parameters and fitted
utilizing TEMPO2. With the best parameters, the timing residuals
distribute near zeros with the root mean squared value 85\,$\mu$s.
However, the timing residuals show slow variations with time
as the Crab pulsar has the large timing noise as shown
in Figure \ref{fig3} (b). In order to verify these results, we also checked
timing residuals observed from Fermi-LAT the at the same time interval
with the same spin parameters. As illustrated in Figure \ref{fig3} (b),
the timing residuals observed by POLAR are remarkably
consistent with Fermi-LAT results, which suggests that timing
system of POLAR is reliable.

\begin{table*}

\footnotesize
\caption {The timing parameters of the Crab pulsar}
\scriptsize{}\label{table1}
\medskip
\begin{center}
\scalebox{1.1}{
\begin{tabular}{l c c }
\hline \hline
& Parameters                            &    Value      \\
\hline \hline
& PEPOCH(MJD)                           & 57697.040344079745    \\
& F0(Hz)                                & 29.6484272934(4)     \\
& F1($10^{-10}$\,Hz\,s$^{-1}$)            & -3.689865(1)      \\
& F2($10^{-20}$\,Hz\,s$^{-2}$)            & 1.16(1)      \\
& F3($10^{-28}$\,Hz\,s$^{-3}$)            & 3.4(3)      \\
\hline
\end{tabular}}
\end{center}
\end{table*}

\subsection{Pulse profile}

\begin{figure}
\begin{center}
\includegraphics[width= 0.9\textwidth, height = 10cm]{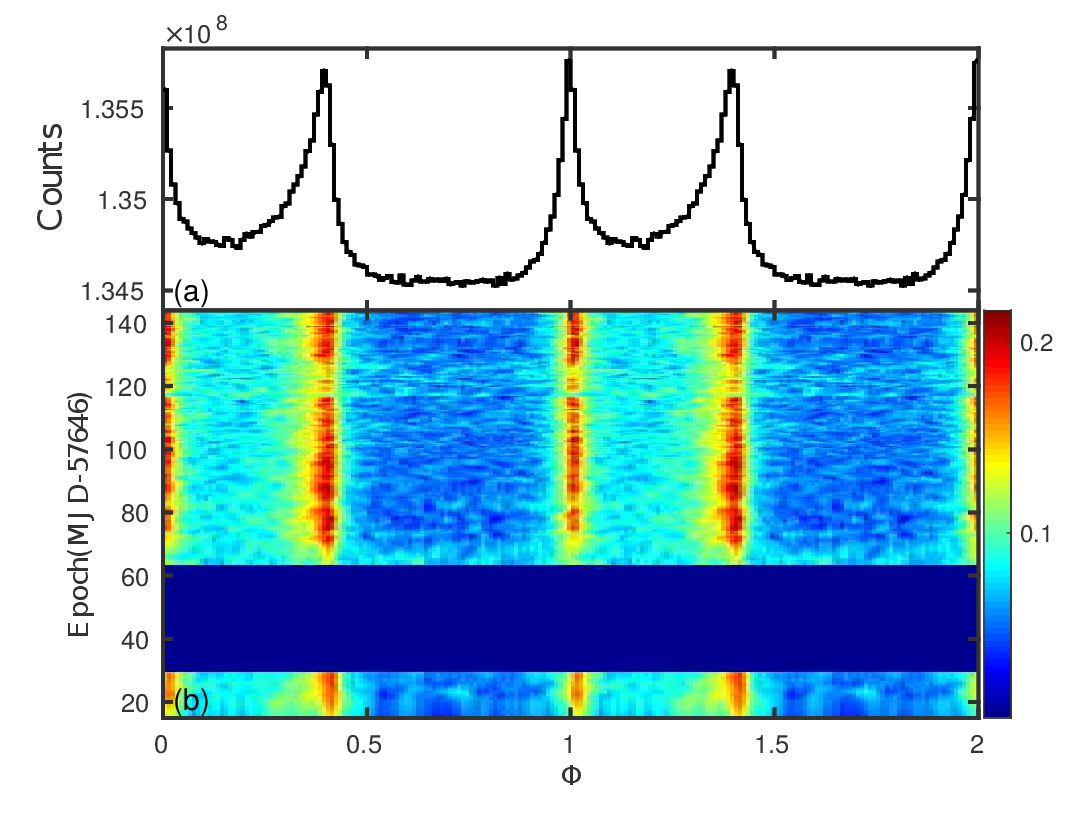}
\caption{The pulse profile detected by POLAR as function of time.
Panel (a) shows the total pulse profile accumulated from all observations.
Panel (b) exhibits the normalized profiles with background subtraction
observed in every day.
\label{fig4}}
\end{center}
\end{figure}

With the accurate timing parameters, the total pulse profile of the Crab pulsar
was folded from the all observed events. As shown in Figure \ref{fig4} (a), the
pulse profile shows the typical double-peak structure with high significance,
which is also consistent with the results of RXTE \cite{Ge(2016)}. As
the large of view of POLAR, Crab could be observed in every orbit and pulse
profile could be obtained though with lower significance. Therefore, all pulse
profiles observed in every day are co-aligned with the same phase as
illustrated in Figure \ref{fig4} (b). These results also confirm that POLAR has
detected the pulsed photons from the compact objects.

\begin{figure}[htbp]
\centering
\subfigure[]{
\begin{minipage}{7cm}
\centering
\includegraphics[width= 7.5 cm,height=7cm]{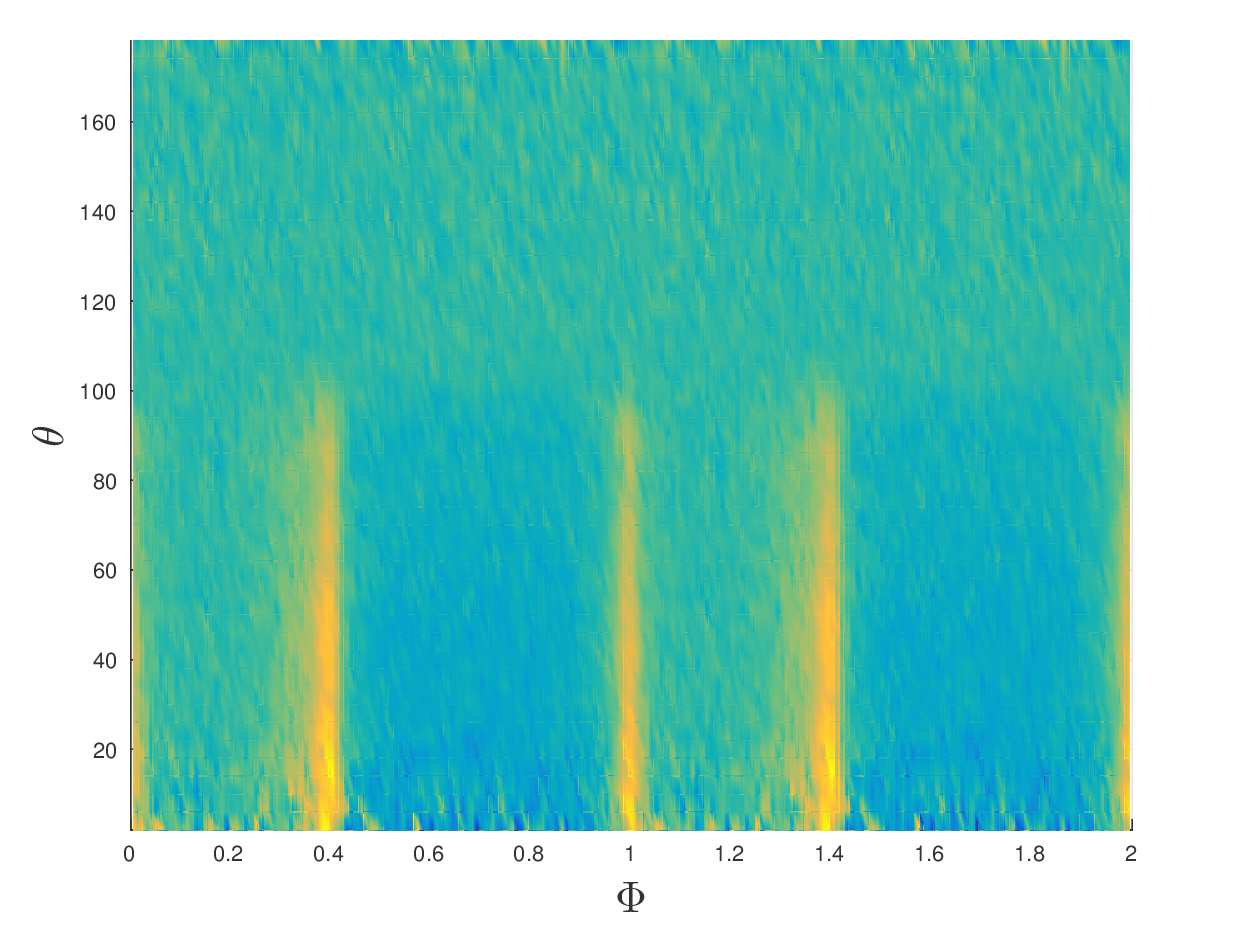}
\end{minipage}
}
\subfigure[]{
\begin{minipage}{7cm}
\centering
    \includegraphics[width= 7.5 cm,height=7cm]{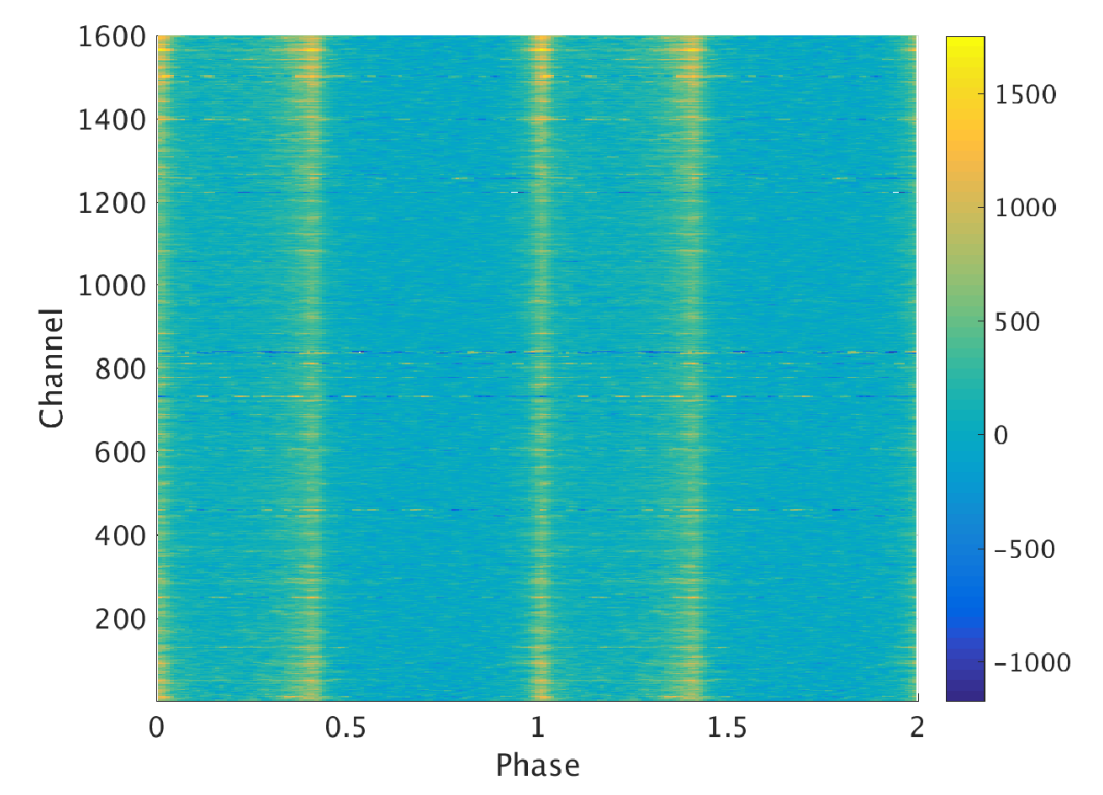}
\end{minipage}
}
\caption{(a) is Pulse profile as function of $\theta$; (b) is pulse profile for 1600 channels.
\label{fig5}}
\end{figure}

The pulse profile as a function of theta incident angle as shown in Figure \ref{fig5} (a),
which indicate that POLAR's FoV in its energy range is larger than 2*$\pi$.
And the pulse profile as a function of 1600 channels as shown in Figure \ref{fig5} (b), which show that the pulsed photons of Crab are captured by every channel.

\subsection{Potential polarization measurement}

All information of an event with its scattered photon (simultaneous trigger) is
recorded by detector. Generally, the first reaction point has maximum deposited
energy, and the second reaction point takes second place. Projecting the direction
from the first reaction point to the second reaction point onto X-Y plane, the
angle between projection and X-axis is the azimuthal angle of this compton
scattering event. For GRB, since it's short duration, accumulating the events in different azimuthal angle, we obtain modulation curve to reveal polarization information. However, for Crab pulsar the incident angle between the Crab and POLAR varies with time, so we have to obtain modulation curve as a function of incident angle. Then, we use POLAR Monte Carlo simulation software package \cite{Kole(2017)} to reappear Crab observation on POLAR. This software package was developed from a calibration experiment on POLAR at European Synchrotron Radiation Facility. Simulating different polarization Crab source in parallel, and comparing these simulation results with observation results. If simulation results of one presuppose polarization are consistent with observation results, then we consider this presuppose polarization as measured polarization of Crab pulsar.

However, the precondition is that we should try to use Monte Carlo simulation to reappear the observation results. In other words, we need a probably calibration process. The response of Crab detection varies with many factors, including but not limited to detection efficiency, projected area, visible efficiency, counts Rate and so on. Look for as many factors as possible from observation-simulation conjoint analysis, and make them decouple with polarization. Then polarization measurement of Crab pulsar is the potential target.

\section{Summary}

Firstly, from the observations, POLAR detected highly significant pulsed signals
from the Crab pulsar which is similar with the known results. The pulse profile as
a function of theta incident angle indicate that POLAR's FoV larger than 2$\pi$,
and as a function of 1600 channels shows that photons of the Crab pulsar were captured by every channel. Second,
the accurate timing behavior observed by POLAR is highly consistent with Fermi-LAT observation,
it verified that POLAR's clock precision was more stable than 85 $\mu$s. The
above results show that POLAR has a good detection performance. And lastly,
we have potential to measure the polarization of Crab on POLAR.

\section*{Acknowledgments}

We thank the High Energy Astrophysics Science Archive Research
Center(HEASARC) at NASA/Goddard Space Flight Center for maintaining its online
archive service that provided the data used in this research. This work is supported
by the National Key Research and Development Program of China (2016YFA0400802), National
Science Foundation of China (11233001, 11503027, 11303069 and 11503028), and the
Strategic Priority Research Program on Space Science, the Chinese Academy of
Sciences,Grant No. XDA04010300 and XDB23000000.


\begin{thebibliography}{99}
\bibitem{Eikenberry(1997)} Eikenberry, S. S., and Fazio, G. G. 1997, \emph{ApJ}, 476, 281
\bibitem{Kuiper(2001)} Kuiper, L., Hermsen, W., Cusumano, G., et al. 2001, \emph{A\&A}, 378, 918
\bibitem{Rots(1998)} Rots, A. H., Jahoda, K., Macomb, D. J., Kawai, N., Saito, et al. \ 1998, \emph{ApJ}, 501, 749
\bibitem{Molkov(2010)} Molkov, S., Jourdain, E., and Roques, J. P.  2010, \emph{ApJ}, 708, 403
\bibitem{Ge(2012)} Ge, M. Y., Lu, F. J., Qu, J. L., Zheng, S. J., Chen, Y. and Han, D. W. 2012, \emph{ApJS}, 199, 32
\bibitem{Novick(1972)} Novick R., et al., 1972, \emph{ApJ}, 174, L1
\bibitem{Weisskopf(1976)} Weisskopf M. C., et al., 1976, \emph{ApJ}, 208, 125
\bibitem{Forot(2008)} M. Forot, et al., \emph{Astrophys. J.} 688 (2008) L29
\bibitem{Dean(2008)} A.J. Dean, et al., \emph{Science} 321 (2008) 1183.
\bibitem{Chauvin(2017)} M. Chauvin, et al. 2017, arXiv:1706.09203 [astro-ph.HE].
\bibitem{Merlin(2017)} M. Kole et al., \emph{ICRC Conf. Proc.} 2017
\bibitem{Sun(2012)} Sun, J. C., Doctoral Thesis, UCAS 2012, at http://ir.ihep.ac.cn/handle/311005/210194
\bibitem{Produit(2005)} N. Produit et al., \emph{Nucl. Instr. and Meth. A} 550 (2005) 616.
\bibitem{Edwards(2006)} Edwards, R.~T., Hobbs, G.~B., \& Manchester, R.~N. 2006, \emph{MNRAS}, 372, 1549
\bibitem{Hobbs(2006)} Hobbs, G.~B., Edwards, R.~T., \& Manchester, R.~N. 2006, \emph{MNRAS}, 369, 655
\bibitem{Lyne(1993)} Lyne, A. G., Pritchard, R. S. and Graham-Smith, F.  1993. \emph{MNRAS}, 265, 1003
\bibitem{Ge(2016)} Ge, M. Y., Yan, L. L., Lu, F. J., et al. 2016, \emph{ApJ}, 818,48
\bibitem{Kole(2017)} M. Kole, Li,Z. H., et al., submitted to \emph{Nucl. Instr. and Meth. A}


\end{thebibliography}
\end{document}